\documentclass[prd,aps,preprintnumbers,amssymb]{revtex4}
\usepackage{axodraw}
\usepackage{color}
\usepackage{epsf}

\def\e3p{$\eta \rightarrow 3 \pi$}

\begin{document}
\title{%
\hfill{\normalsize\vbox{%
\hbox{}
 }}\\
{  Non-perturbative renormalization of the Yang-Mills  beta functions}}

\author{Renata Jora
$^{\it \bf a}$~\footnote[2]{Email:
 rjora@theory.nipne.ro}}

\author{Salah Nasri
 $^{\it \bf b, \bf c}$~\footnote[4]{Email:
 nasri.salah@gmail.com}}

\affiliation{$^{\bf \it a}$ National Institute of Physics and Nuclear Engineering PO Box MG-6, Bucharest-Magurele, Romania}

\affiliation{$^ {\bf \it b}$ Department of Physics,
 United Arab Emirates University, UAE}

\affiliation{$^ {\bf \it c}$International Center for Theoretical Physics, Trieste, Italy}

\date{\today}

\begin{abstract}
We determine the non-perturbative corrections to the gauge coupling constant and the topological charge in the Yang Mills theory. The method makes no explicit use of instanton calculations but instead  relies on boundary properties of the quantum partition function. The approach may offer important clues regarding the behavior of the coupling constant and the theta angle for other gauge theories like QCD and supersymmetric QCD.

\end{abstract}

\maketitle

\section{Introduction}

The effective Lagrangian for a gauge theory is characterized by both perturbative and non-perturbative contributions. Peturbative corrections by themselves are  easier to find because they are based on repetitive physical and mathematical algorithms. For the non-perturbative terms one needs to calculate the various instanton contributions.  Important results regarding the  effective action for supersymmetric QCD theories were obtained in \cite{Seiberg1},\cite{Seiberg2}. In \cite{Seiberg3} the non-perturbative beta function of the supersymmetric $N=2$ gauge theories was estimated.  For Yang Mills and QCD like theories the non-perturbative beta functions were calculated through instanton terms. In \cite{Callan} the non-perturbative contribution to the Yang Mills beta function of the coupling constant was determined. Later on in \cite{Morozov} both non-perturbative contributions to the beta function of the coupling constant and that of the topological charge were reobtained in good agreement with the previous results. These are:
\begin{eqnarray}
&&\frac{1}{g^2(\mu')}=\frac{1}{g^2(\mu)}+\frac{b_0}{8\pi^2}\ln[\frac{\mu}{\mu}]-4\ln[\frac{\mu}{\mu'}]D(\mu)\cos(\theta(\mu))
\nonumber\\
&&\theta (\mu')=\theta(\mu)-32\pi^2\ln[\frac{\mu}{\mu'}]D(\mu)\sin(\theta(\mu)).
\label{firstcontr665}
\end{eqnarray}
Here $D(\mu)$ is a power function of $\mu$, $g^2(\mu)$ is the coupling constant at scale $\mu$ and $g^2(\mu')$ is the coupling constant at scale $\mu'$. Furthermore $b_0$ is the first order of the perturbative gauge coupling beta function.

The non-perturbative beta functions in QCD were also calculated in \cite{Jora} with results that differ slightly from those in \cite{Morozov}.

In this paper we will determine the non-perturbative contribution to the gauge coupling constant and the topological charge in a novel method in which instanton transitions are taken into account without computing explicitly any instanton but by estimating them from the boundary behavior of the unitary operators that appear in a quantum calculations.  We will show that this kind of procedure which is clean and straightforward can replace with success the more intricate and messy instanton calculations.

Section II contains the set-up and the general method. Section III includes explicit calculations of the non-perturbative contributions to the beta functions of the gauge coupling constant and that of the theta vacuum.  The last section is dedicated to the conclusions.

\section{The set-up}

We start with the Yang Mills Lagrangian in the presence of a $\theta$ term.
\begin{eqnarray}
{\cal L}=-\frac{1}{4g^2}F^{a\mu\nu}F^a_{\mu\nu}-i\theta\frac{1}{32\pi^2}\tilde{F}^{a\mu\nu}F^a_{\mu\nu},
\label{lag455344389}
\end{eqnarray}
where $\tilde{F}^{a\mu\nu}=\frac{1}{2}\epsilon^{\mu\nu\rho\sigma}F^a_{\rho\sigma}$. We observe that one can write:
\begin{eqnarray}
{\cal L}(X)=-\frac{1}{8g^2}X^{\mu\nu\rho\sigma}X_{\mu\nu\alpha\beta}F^a_{\rho\sigma}F^{a\alpha\beta}-i\frac{\theta}{64\pi^2}X^{\mu\nu\alpha\beta}F^a_{\alpha\beta}F^a_{\mu\nu},
\label{secondofmrlagr4536}
\end{eqnarray}
provided that:
\begin{eqnarray}
X^{\mu\nu\alpha\beta}=\epsilon^{\mu\nu\alpha\beta}.
\label{deftensor7564}
\end{eqnarray}
Then $X^{\mu\nu\alpha\beta}$ may be regarded a general rank four tensor which may be space time dependent. Let us consider the following change in this tensor:
\begin{eqnarray}
X^{\prime \mu\nu\alpha\beta}=X^{\mu\nu\alpha\beta}+\alpha X_{\rho\sigma}^{\alpha\beta}\epsilon^{\rho\sigma\mu\nu},
\label{change546354}
\end{eqnarray}
where $\alpha$ is an infinitesimal parameter. The Lagrangian in Eq. (\ref{secondofmrlagr4536}) will then be modified (with the condition in Eq. (\ref{deftensor7564})) to:
\begin{eqnarray}
{\cal L}'={\cal L}-\alpha\Bigg[\frac{1}{g^2}F^a_{\mu\nu}\tilde{F}^{a\mu\nu}+i\theta\frac{1}{32\pi^2}F^a_{\mu\nu}F^a_{\mu\nu}\Bigg].
\label{transf453635}
\end{eqnarray}
Note that the main effect is the transformation of the gauge tensor into its dual.

Consider the effective Lagrangian ${\cal L(\mu}$ \cite{Morozov} obtained by integrating our the perturbative fluctuations with momenta $p^2>\mu^2$ and over instantons with a size $\rho<\frac{1}{\mu}$:
\begin{eqnarray}
{\cal L} (\mu)=\frac{1}{4g^2(\mu)}F^{a\mu\nu}F^a_{\mu\nu}+i\theta(\mu)\frac{1}{32\pi^2}\tilde{F}^{a\mu\nu}F^a_{\mu\nu}.
\label{res534288}
\end{eqnarray}
The effective action at scale $\mu'<\mu$ is obtained by integrating out the fluctuations between $\mu$ and $\mu'$. Therefore:
\begin{eqnarray}
&&Z(\mu')=\exp\Bigg[\int d^4x [\frac{1}{4g^2(\mu')}F^{a\mu\nu}F^a_{\mu\nu}+i\theta(\mu')\frac{1}{32\pi^2}\tilde{F}^{a\mu\nu}F^a_{\mu\nu}]\Bigg]=
\nonumber\\
&&\int dA^a_{\mu}\exp[\int d^4x {\cal L}(\mu,B^a_{\nu},A^a_{\nu})],
\label{expr6453424}
\end{eqnarray}
where $B^a_{\nu}$ are the fields with momenta between $0$ and $\mu'$ and $A^a_{\mu}$ are the fluctuations with momenta in the range $[\mu',\mu]$. Thus $F^a_{\mu\nu}$ in the first line of Eq. (\ref{expr6453424}) is expressed entirely in terms of $B^a_{\mu}$.

One can write:
\begin{eqnarray}
&&Z(\mu')=\exp\Bigg[\int d^4x [\frac{1}{4g^2(\mu')}F^{a\mu\nu}F^a_{\mu\nu}+i\theta(\mu')\frac{1}{32\pi^2}\tilde{F}^{a\mu\nu}F^a_{\mu\nu}]\Bigg]=
\nonumber\\
&&\int dA^a_{\mu}\int d X^{\mu\nu\alpha\beta} \delta(X^{\mu\nu\alpha\beta}-\epsilon^{\mu\nu\alpha\beta})\exp[\int d^4x {\cal L}(\mu,B^a_{\nu},A^a_{\nu})],
\label{expr6453424new}
\end{eqnarray}
where the tensor $X^{\mu\nu\alpha\beta}$ is introduced in ${\cal L}(\mu)$ according to the same prescription as in Eq. (\ref{secondofmrlagr4536}). Then an infinitesimal change in $X^{\mu\nu\alpha\beta}$ as in (\ref{change546354}) should not affect the integral. Accordingly,
\begin{eqnarray}
&&\int dA^a_{\mu}\int d X^{\mu\nu\alpha\beta}\delta(X^{\mu\nu\alpha\beta}-\epsilon^{\mu\nu\alpha\beta}) \exp[\int d^4x {\cal L}(X,B^a_{\nu},A^a_{\nu})]=
\nonumber\\
&&\int dA^a_{\mu}\int d X^{\prime\mu\nu\alpha\beta} \delta(X^{\prime\mu\nu\alpha\beta}-\epsilon^{\mu\nu\alpha\beta})\exp[\int d^4x {\cal L}(X',B^a_{\nu},A^a_{\nu})].
\label{res645342}
\end{eqnarray}
The variation proportional to $\alpha$ in the second line of Eq. (\ref{res645342}) can come form three sources: the exponential of the action, the delta function and the jacobian of the variable of integration.

Let us first discuss the contribution from the jacobian:
\begin{eqnarray}
&&d X^{\prime \mu\nu \alpha\beta}=\frac{d X^{\prime \mu\nu \alpha\beta}}{d X^{\rho\sigma\gamma\delta}}d X^{\rho\sigma\gamma\delta}=
\nonumber\\
&&\det[\delta^{\mu}_{\rho}\delta^{\nu}_{\sigma}\delta^{\alpha}_{\gamma}\delta^{\beta}_{\delta}+\alpha\delta^{\alpha}_{\gamma}\delta^{\beta}_{\delta}\epsilon^{\mu\nu\rho\sigma}]d X^{\rho\sigma\gamma\delta}=
\nonumber\\
&&\exp\Bigg[{\rm Tr} \ln[\delta^{\mu}_{\rho}\delta^{\nu}_{\sigma}\delta^{\alpha}_{\gamma}\delta^{\beta}_{\delta}+\alpha\delta^{\alpha}_{\gamma}\delta^{\beta}_{\delta}\epsilon^{\mu\nu\rho\sigma}]\Bigg]
d X^{\rho\sigma\gamma\delta}=
\nonumber\\
&&\exp\Bigg[{\rm Tr}[\alpha\delta^{\alpha}_{\gamma}\delta^{\beta}_{\delta}\epsilon^{\mu\nu\rho\sigma}]\Bigg]d X^{\rho\sigma\gamma\delta}=d X^{\rho\sigma\gamma\delta}.
\label{jaconco56}
\end{eqnarray}
It is clear that the jacobian brings no contribution proportional to the small parameter $\alpha$.

The next step is to consider the contribution of the delta function:
\begin{eqnarray}
&&\frac{\partial\prod\delta(X^{\prime\mu\nu\alpha\beta}-\epsilon^{\mu\nu\alpha\beta})}{\partial \alpha}=
\nonumber\\
&&\prod\delta(X^{\mu\nu\alpha\beta}-\epsilon^{\mu\nu\alpha\beta}){\rm Tr}\Bigg[-\epsilon^{\mu\nu\gamma\delta}X^{\alpha\beta}_{\gamma\delta}\frac{\partial }{\partial X^{\mu\nu\alpha\beta}}\Bigg].
\label{deltacontr65788}
\end{eqnarray}
Introduced in the partition function Eq. (\ref{deltacontr65788}) leads to:
\begin{eqnarray}
&&(\delta Z(\mu'))_1=\frac{\partial }{\partial \alpha}\int dA^a_{\mu}\int d X^{\mu\nu\alpha\beta}\prod \delta(X^{\prime\mu\nu\alpha\beta}-\epsilon^{\mu\nu\alpha\beta})\exp[\int d^4x {\cal L}(X,B^a_{\nu},A^a_{\nu}]=
\nonumber\\
&&\int d X^{\mu\nu\alpha\beta}\prod\delta(X^{\mu\nu\alpha\beta}-\epsilon^{\mu\nu\alpha\beta}){\rm Tr}\Bigg[-\epsilon^{\mu\nu\gamma\delta}X^{\alpha\beta}_{\gamma\delta}\frac{\partial }{\partial X^{\mu\nu\alpha\beta}}\Bigg]\int dA^a_{\mu}\exp[\int d^4x {\cal L}(X,B^a_{\nu},A^a_{\nu}]=
\nonumber\\
&&{\rm Tr}\Bigg[\int d X^{\mu\nu\alpha\beta}\delta(X^{\mu\nu\alpha\beta}-\epsilon^{\mu\nu\alpha\beta})[-\delta^{\mu\alpha}\delta^{\nu\beta}\frac{\partial }{\partial X^{\mu\nu\alpha\beta}}]\int dA^a_{\mu}\exp[\int d^4x {\cal L}(X,B^a_{\nu},A^a_{\nu}]\Bigg]=0.
\label{res645553}
\end{eqnarray}
where in the last line all delta functions were integrated except that remaining in the trace.

The last contribution is that coming form the exponential of the action:
\begin{eqnarray}
&&(\delta  Z(\mu'))_2= \int dA^a_{\mu}\Bigg[\int d^4 x[\frac{1}{g^2(\mu)}\tilde{F}^{a\mu\nu}(B,A)F^a_{\mu\nu}(B,A)+i\theta(\mu)\frac{1}{32\pi^2}F^{a\mu\nu}(B,A)F^a_{\mu\nu}(B,A)]\Bigg]\times
\nonumber\\
&&\exp[\int d^4x {\cal L}(\mu,B^a_{\nu},A^a_{\nu})].
\label{finalres6455369}
\end{eqnarray}
Here we used Eq. (\ref{transf453635}) for the variation of the Lagrangian with respect to $\alpha$. Note also that the gauge tensor depend on both $B^a_{\mu}$ with momenta smaller than $\mu'$ and $A^a_{\mu}$ with momenta between $\mu'$ and $\mu$.

Finally since  a change in the integration variable should not affect the full result of the infinite integral we have:
\begin{eqnarray}
(\delta Z)_1+(\delta Z)_2=0
\label{finalres55454}
\end{eqnarray}

\section{Beta functions for the coupling constant and for the theta vacuum}

We start with the expression in Eq. (\ref{finalres6455369}):
\begin{eqnarray}
&&(\delta  Z(\mu'))_2= \int dA^a_{\mu}\Bigg[\int d^4x[\frac{1}{g^2(\mu)}\tilde{F}^{a\mu\nu}(B,A)F^a_{\mu\nu}(B,A)+i\theta\frac{1}{32\pi^2}F^{a\mu\nu}(B,A)F^a_{\mu\nu}(B,A)]\Bigg]\exp[\int d^4x {\cal L}(\mu,B^a_{\nu},A^a_{\nu})]=
\nonumber\\
&&\int dA^a_{\mu}\Bigg[\frac{1}{g^2(\mu)}\frac{\partial }{\partial [i\theta\frac{1}{32\pi^2}]}+i\theta\frac{1}{32\pi^2}\frac{\partial}{\partial \frac{1}{g^2}}\Bigg]\exp[\int d^4x {\cal L}(\mu,B^a_{\nu},A^a_{\nu})]=
\nonumber\\
&&\int dA^a_{\mu}\Bigg[-i\frac{32\pi^2}{g^2}\frac{\partial}{\partial \theta}+i\theta\frac{1}{8\pi^2}\frac{\partial}{\partial \frac{1}{g^2}}\Bigg]\exp[\int d^4x {\cal L}(\mu,B^a_{\nu},A^a_{\nu})].
\label{firsteq42332}
\end{eqnarray}
Here the exponent in all the lines  of the equation is just given by the expression in Eq.(\ref{res534288}). A derivative with finite limits of integration affects both the integrand and the limits of integration. Let us rewrite in interaction picture the effective actions we are dealing with to have  a clearer picture.
\begin{eqnarray}
&&Z(\mu)=\langle 0|U(\mu,-T)|0\rangle
\nonumber\\
&&Z(\mu')=\langle 0|U(\mu',-T)|0\rangle=\langle \mu|\langle 0|U(\mu,-T)|0\rangle|\mu'\rangle,
\label{res54663}
\end{eqnarray}
where $U$ is the unitary evolution operator.
Practically in the last line of Eq. (\ref{firsteq42332}) we have a derivative applied to $\langle 0|U(-T,\mu)|0\rangle$. Let us write:
\begin{eqnarray}
\Bigg[-i\frac{32\pi^2}{g^2}\frac{\partial}{\partial \theta}+i\theta\frac{1}{8\pi^2}\frac{\partial}{\partial \frac{1}{g^2}}\Bigg]=W(\mu)\frac{d}{d(\mu)}.
\label{defuseful564788}
\end{eqnarray}
Then,
\begin{eqnarray}
&&\langle \mu|\frac{d}{d \mu}[\langle 0|U(\mu,-T)|0\rangle]|\mu'\rangle=
\nonumber\\
&&\frac{d}{d \mu}\Bigg[\langle \mu|\langle 0|U(\mu,-T)|0\rangle|\mu'\rangle\Bigg]-
\nonumber\\
&&\frac{d}{d \mu}[\langle \mu|]\langle 0|U(\mu,-T)|0\rangle|\mu'\rangle-\langle \mu|\langle 0|U(\mu,-T)|0\rangle|\frac{d}{d \mu}[\mu'\rangle],
\label{res55342}
\end{eqnarray}
so the derivative of the integrand is the full derivative of the integral minus the derivatives applied to the limits of integration. Knowing that:
\begin{eqnarray}
&&\langle \mu|=\langle 0|U(T,\mu)
\nonumber\\
&&|\mu'\rangle=U(\mu',-T)|0\rangle,
\label{evol647736}
\end{eqnarray}
one can write,
\begin{eqnarray}
&&\frac{d}{d \mu}[\langle \mu|]=-{\cal L}_1(\mu)\langle \mu|
\nonumber\\
&&\frac{d}{d \mu}[\mu'\rangle]=\int d \mu'\delta(\mu'-\mu){\cal L}(\mu')|0\rangle.
\label{somecalc756478}
\end{eqnarray}
Here we took into account the order in the limits of integration and ${\cal L}_1(\mu)=-{\cal L}(\mu)$.

Finally introducing the results of Eq, (\ref{evol647736}) into Eq. (\ref{res55342}) one obtains:
\begin{eqnarray}
&&-\frac{d}{d \mu}[\langle \mu|]\langle 0|U(\mu,-T)|0\rangle|\mu'\rangle-\langle \mu|\langle 0|U(\mu,-T)|0\rangle|\frac{d}{d \mu}[\mu'\rangle]=
\nonumber\\
&&\langle \mu|{\cal L}_1(\mu)\langle 0|U(\mu,-T)|0\rangle]|\mu'\rangle-\langle \mu|\langle 0|U(\mu,-T)|0\rangle]{\cal L}(\mu)|0\rangle.
\label{finishcomecalc7564884}
\end{eqnarray}
Furthermore,
\begin{eqnarray}
&&\langle \mu|{\cal L}_1(\mu)\langle 0|U(\mu,-T)|0\rangle]|\mu'\rangle=
\nonumber\\
&&\int dA^a_{\mu}\Bigg[\int d^4x[\frac{1}{g^2(\mu)}\tilde{F}^{a\mu\nu}(B)F^a_{\mu\nu}(B)+i\theta\frac{1}{32\pi^2}F^{a\mu\nu}(B)F^a_{\mu\nu}(B)]\Bigg]\exp[\int d^4x {\cal L}(\mu,B^a_{\nu},A^a_{\nu}],
\label{psme66455}
\end{eqnarray}
equation which is obtained by applying the operator ${\cal L}_1(\mu)$ or equivalently the corresponding hamiltonian of interaction to the right state.

Retrieving the initial operators an taking into account that,
\begin{eqnarray}
\langle \mu|\langle 0|U(-T,\mu)|0\rangle]|0\rangle=z,
\label{domesimpl6776}
\end{eqnarray}
leads to (Here $z$ is a constant related to the normalization of the partition function):
\begin{eqnarray}
&&(\delta  Z(\mu'))_2=\Bigg[-i\frac{32\pi^2}{g^2}\frac{\partial}{\partial \theta}+i\theta\frac{1}{8\pi^2}\frac{\partial}{\partial \frac{1}{g^2(\mu)}}\Bigg]\times\exp[\int d^4 x{\cal L}(\mu',B)]+
\nonumber\\
&&\int dA^a_{\mu}\Bigg[\int d^4x[-\frac{1}{g^2(\mu)}\tilde{F}^{a\mu\nu}(B)F^a_{\mu\nu}(B)-i\theta\frac{1}{32\pi^2}F^{a\mu\nu}(B)F^a_{\mu\nu}(B)]\Bigg]\exp[\int d^4x {\cal L}(\mu,B^a_{\nu},A^a_{\nu}]-
\nonumber\\
&&z\Bigg[\int d^4 x[\frac{1}{g^2(\mu)}\tilde{F}^{a\mu\nu}(B)F^a_{\mu\nu}(B)+i\theta\frac{1}{32\pi^2}F^{a\mu\nu}(B)F^a_{\mu\nu}(B)]\Bigg].
\label{finalres546635}
\end{eqnarray}
Note that on the third line there is no exponential and the tensor fields are expressed only in terms of $B$ as they stem from the lagrangian at the scale $\mu'$.

Furthermore we observe that,
\begin{eqnarray}
&&(\delta  Z(\mu'))_1=0,
\label{onee3554}
\end{eqnarray}
which further leads to:
\begin{eqnarray}
&&(\delta  Z(\mu'))_1+(\delta  Z(\mu'))_2=\Bigg[-i\frac{32\pi^2}{g^2}\frac{\partial}{\partial \theta}+i\theta\frac{1}{8\pi^2}\frac{\partial}{\partial \frac{1}{g^2}}\Bigg]\times\exp[\int d^4 x{\cal L}(\mu',B)]+
\nonumber\\
&&\Bigg[\int d^4x[-\frac{1}{g^2(\mu)}\tilde{F}^{a\mu\nu}(B)F^a_{\mu\nu}(B)-i\theta(\mu)\frac{1}{32\pi^2}F^{a\mu\nu}(B)F^a_{\mu\nu}(B)]\Bigg]\times\exp[\int d^4 x{\cal L}(\mu',B)]+
\nonumber\\
&&z\int d^4x[\frac{1}{g^2(\mu)}\tilde{F}^{a\mu\nu}(B)F^a_{\mu\nu}(B)+i\theta(\mu)\frac{1}{32\pi^2}F^{a\mu\nu}(B)F^a_{\mu\nu}(B)]=0.
\label{thisisrealfinal65748}
\end{eqnarray}

We proceed further to calculate and simplify Eq. (\ref{thisisrealfinal65748}) to obtain:
\begin{eqnarray}
&&\Bigg[-i\frac{8\pi^2}{g^2}\frac{\partial \frac{1}{g^{\prime2}}}{\partial \theta}+i\theta\frac{1}{32\pi^2}\frac{\partial \frac{1}{g^{\prime 2}}}{\partial \frac{1}{g^2}}\Bigg]FF+
\nonumber\\
&&\Bigg[\frac{1}{g^2}\frac{\partial \theta'}{\partial \theta}-\theta\frac{1}{(8\pi^2)(32\pi^2)}\frac{\partial \theta'}{\partial \frac{1}{g^2}}\Bigg]F\tilde{F}+
\nonumber\\
&&\Bigg[-\frac{1}{g^2}F\tilde{F}-i\theta\frac{1}{32\pi^2}FF\Bigg]=
\nonumber\\
&&z\Bigg[\frac{1}{g^2}F\tilde{F}+i\theta\frac{1}{32\pi^2}FF\Bigg]\exp[-d^4x{\cal L}(B,\mu')].
\label{tosee3443}
\end{eqnarray}
Here we used the notations: $g^2(\mu)=g^2$, $g^2(\mu')=g^{\prime 2}$, $\tilde{F}^{a\mu\nu}(B)F^a_{\mu\nu}(B)=F\tilde{F}$ and $F^{a\mu\nu}(B)F^a_{\mu\nu}(B)=FF$

In the background of an instanton with winding number $n=1$ and by integrating the full equation in (\ref{tosee3443}) one obtains:
\begin{eqnarray}
&&-i\frac{1}{g^2}\frac{\delta [\frac{8\pi^2}{g^{\prime 2}}+i\theta']}{\partial \theta}+i\theta\frac{1}{(8\pi^2)(32\pi^2)}\frac{\delta [\frac{8\pi^2}{g^{\prime 2}}+i\theta']}{\partial \frac{1}{g^2}}-\frac{1}{g^2}-i\theta\frac{1}{32\pi^2}=
\nonumber\\
&&-z\Bigg[\frac{1}{g^2}+i\theta\frac{1}{32\pi^2}\Bigg]\exp[-\frac{8\pi^2}{g^{\prime2}}-i\theta'].
\label{finalres664554}
\end{eqnarray}
We will first solve the equation with a few simplifying assumptions. One is to  consider that the perturbative beta function runs at one loop, the second is to neglect  non-perturbative corrections to the instanton in the exponential. Then one finds:
\begin{eqnarray}
\frac{8\pi^2}{g^{\prime 2}}+i\theta'=\frac{8\pi^2}{g^2}+\beta_0\ln[\frac{\mu'}{\mu}]+z[\frac{\mu}{\mu'}]^{\beta_0}\exp[-\frac{8\pi^2}{g^2}-i\theta].
\label{res6645530}
\end{eqnarray}
Here the one loop perturbative beta function of the Yang-Mills theory is:
\begin{eqnarray}
\beta(g)=-\beta_0\frac{1}{16\pi^2}g^3.
\label{betafunctr66565}
\end{eqnarray}

Next we will find a general solution to Eq. (\ref{finalres664554}). We denote:
\begin{eqnarray}
\frac{1}{g^{\prime 2}}+i\theta'=\frac{1}{g^2}+i\theta+f,
\label{not657465}
\end{eqnarray}
where $f$ contains both perturbative and non-perturbative corrections.  Then Eq. (\ref{finalres664554}) simplifies to:
\begin{eqnarray}
&&\frac{1}{g^2}\frac{\partial f}{\partial (i\theta)}+i\theta\frac{1}{32\pi^2}\frac{\partial f}{\partial (\frac{8\pi^2}{g^2})}=
\nonumber\\
&&-z[\frac{1}{g^2}+i\theta\frac{1}{32\pi^2}]\exp[-\frac{8\pi^2}{g^2}-i\theta]\exp[-f].
\label{scerround64553}
\end{eqnarray}
This may be rewritten as:
\begin{eqnarray}
&&\frac{1}{g^2}\frac{\partial F}{\partial (i\theta)}+i\theta\frac{1}{32\pi^2}\frac{\partial F}{\partial (\frac{8\pi^2}{g^2})}=
\nonumber\\
&&-z[\frac{1}{g^2}+i\theta\frac{1}{32\pi^2}]\exp[-\frac{8\pi^2}{g^2}-i\theta],
\label{third645546}
\end{eqnarray}
where $F=\exp[f]$. The general solution is then:
\begin{eqnarray}F=c+z\exp[-\frac{8\pi^2}{g^2}-i\theta],
\label{gensol675}
\end{eqnarray}
which further leads to (Here $c$ is a constant of integration):
\begin{eqnarray}
f=\ln\Bigg[c+z\exp[-\frac{8\pi^2}{g^2}-i\theta]\Bigg].
\label{final657774}
\end{eqnarray}
Using Eq. (\ref{not657465}) and by expanding the logarithm in Eq. (\ref{final657774}) one arrives to the final result:
\begin{eqnarray}
\frac{8\pi^2}{g^{\prime 2}}+i\theta'=\frac{1}{g^2}+i\theta+\ln[c]+\frac{z}{c}\exp[-\frac{8\pi^2}{g^2}-i\theta]+...
\label{bettwo887}
\end{eqnarray}
By setting $\ln[c]=[\frac{\mu'}{\mu}]^{\beta_0}$ one retrieves the first order result in Eq. (\ref{betafunctr66565}).

\section{Conclusions}

Although we started from Yang-Mills theories the results in Eqs.(\ref{betafunctr66565}) and (\ref{bettwo887}) are in complete agreement with the results for the  holomorphic coupling constant of supersymmetric QCD \cite{Seiberg1}, \cite{Seiberg2} obtained through a completely different method. The results are also very similar to previous determinations of the non-perturbative beta functions for Yang-Mills theories as given
in Eq. (\ref{firstcontr665}).  The main difference is the absence of the logarithmic factor in front of the non-perturbative term. But this may be easily accounted for by the constant $z$ which is connected to the exact normalization of the partition function.

We considered the $n=1$ instanton background but the beta function contains infinite instanton contributions with coefficients in front that are related to one another in a simple way.  Thus our result may represent a significant advance in disentangling the non-perturbative contributions for any gauge theory.


\begin{thebibliography}{15}
\bibitem{Seiberg1} N. Seiberg, Nucl. Phys. B {\bf 435}, 129 (1995).
\bibitem{Seiberg2} K. A. Intriligator and N. Seiberg, Nucl. Phys. Proc. Suppl. {\bf 45} BC:1-28 (1996).
\bibitem{Seiberg3} N. Seiberg, Phys. Lett. B {\bf 206}, 75-80 (1988).
\bibitem{Callan} C. G. Callan, R. Dashen and D. J. Gross, Phys. Rev. D {\bf 17}, 2717 (1978).
\bibitem{Morozov} V. G. Knizhnik and A. Yu. Morozov, Pis'ma Zh. Eksp. Teor. Fiz. {\bf 39}, No. 5, 2020205 (1984).
\bibitem{Jora} R. Jora, arXiv:1704.01714 (2017).
\end{thebibliography}
\end{document}